\documentclass[12pt]{iopart}
\usepackage[T1]{fontenc}
\usepackage{graphicx}
\usepackage{hyperref}
\usepackage{soul}
\usepackage{color}

\begin{document}

\title{Initial Stages of FeO Growth on Ru(0001)}

\author{I. Palacio$^{1,2}$\footnote{Present Address: Instituto de Ciencia de Materiales de Madrid (CSIC), Cantoblanco, 28049 Madrid, Spain}, M. Monti$^3$, J.F. Marco$^3$, K.F. McCarty$^4$, J. de la Figuera$^3$}
\address{1 Departamento de F\'{\i}sica de Materiales, Universidad Complutense de Madrid, Madrid 28040 Spain}
\address{2 Unidad Asociada IQFR(CSIC)-UCM, Madrid 28040, Spain}
\address{3 Instituto de Qu\'{\i}mica-F\'{\i}sica ``Rocasolano'', CSIC, Madrid 28006, Spain}
\address{4 Sandia National Laboratories, Livermore, California 94550, USA}

\begin{abstract}

We study how FeO w\"{u}stite films on Ru(0001) grow by oxygen-assisted molecular beam epitaxy at elevated temperatures (800--900~K). The nucleation and growth of FeO islands are observed in real time by low-energy electron microscopy (LEEM). When the growth is performed in an oxygen pressure of 10$^{-6}$~Torr, the islands are of bilayer thickness (Fe-O-Fe-O). In contrast, under a pressure of 10$^{-8}$~Torr, the islands are a single FeO layer thick. We propose that the film thickness is controlled by the concentration of oxygen adsorbed on the Ru. More specifically, when monolayer growth increases the adsorbed oxygen concentration above a limiting value, its growth is suppressed. Increasing the temperature at a fixed oxygen pressure decreases the density of FeO islands. However, the nucleation density is not a monotonic function of oxygen pressure. 

\end{abstract}

\date{\today}

\maketitle

\section{Introduction}
 
Noble metals on FeO surfaces exhibit the well-known ``strong metal support interaction'' (SMSI) in catalysis~\cite{meyer_two-dimensional_2003, boffa_promotion_1994}. Furthermore, different groups have recently demonstrated experimentally~\cite{sun_monolayer_2009,fu_interface-confined_2010} and theoretically~\cite{sun_interplay_2010} how an ultrathin FeO film grown on a noble metal surface presents an enhanced catalytic activity. In particular {FeO$_x$/Pt} promotes CO oxidation and other reactions~\cite{sun_interplay_2010, merte_co-induced_2011, xu_hydroxyls-involved_2011, tanaka_new_2004, sun_when_2008, lei_co+no_2011}. 

Iron binary oxides can adopt several structures, including FeO, Fe$_3$O$_4$, $\gamma$-Fe$_2$O$_3$, and $\alpha$-Fe$_2$O$_3$. Bulk FeO (w\"ustite) crystallizes in the cubic NaCl structure~\cite{CornellBook}. Along the (111) direction, it consists of alternating layers of Fe$^{2+}$ cations and  O$^{2-}$ anions, each arranged in an hexagonal lattice. The distance between adjacent O or Fe atoms in each hexagonal plane is 0.30~nm. FeO is an antiferromagnet with a N\'{e}el temperature of 200~K. The Fe atoms within each (111) plane are coupled ferromagnetically and the different Fe planes are coupled
 antiferromagnetically. The stability of FeO appears to be enhanced on metal substrates such as Pt~\cite{ranke_crystal_1999,weiss_surface_2002} and, in particular, on Ru(0001)~\cite{ketteler_heteroepitaxial_2003}, where several layers can be grown.
 
Ultrathin iron oxides on metal substrates have usually been grown by the surface science community using separate steps of depositing an ultrathin iron film at room temperature followed by oxidation at $\sim$900~K in a molecular oxygen atmosphere. This sequence is repeated if thicker films are desired~\cite{weiss_surface_2002,ketteler_heteroepitaxial_2003,SalaPRB2012,SpiridisPRB2012}. An alternative growth method is that based on oxygen-assisted molecular beam epitaxy (O-MBE), where Fe is deposited in a background pressure of oxygen. Thus the surface is covered with oxygen before Fe arrives at the substrate. Recent work has found that the iron oxide films can be strongly affected by small changes in growth conditions, producing for example either Fe$_3$O$_4$ or FeO~\cite{SpiridisPRB2012}. In the related cobalt oxides, even the crystallographic orientation of an oxide thin film can be selected~\cite{HeinzPRL2012}. Natural questions are then whether the two methods produce the same structures and morphologies and what factors, such as oxygen pressure and temperature, control FeO film growth.

In this work we begin by characterizing O-MBE films on Ru(0001) and show that they yield essentially the same FeO phase as produced by oxidizing Fe films~\cite{ketteler_heteroepitaxial_2003}. However, we find that the O-MBE films can be bilayer in height, unlike the monolayer islands typically produced at the initial stages of Fe oxidation. To understand why FeO bilayers or monolayers are produced, we use low-energy electron microscopy (LEEM~\cite{santos_structure_2009,monti_magnetism_2012, SalaPRB2012}) to image in real space the surface during growth. We propose that the oxygen coverage on the Ru substrate controls whether single layer or bilayer films result. And by analyzing the temperature dependence of the FeO island nucleation density, we estimate the activation energy for surface diffusion of the Fe-O growth species.

\section{Experimental}

The experiments were carried out in two experimental stations. The first is a commercial Elmitec III LEEM microscope. The second is a multipurpose ultra-high vacuum chamber equipped with a home-made scanning tunneling microscope~\cite{DiaconescuRSI2007}, an hemispherical analyzer for X-ray photoelectron spectroscopy and a conventional low-energy electron diffractometer (STM/XPS/LEED system). The XPS system comprises a Al/Mg K$\alpha$ x-ray source and a Phoibos-150 hemispherical analyzer. The spectra were acquired with a pass energy of 20~eV and calibrated relative to the Ru 3$d_\frac{5}{2}$ core level. The STM/XPS/LEED system provides a complementary view of the surface morphology by STM, chemical information by XPS and laterally averaged low-energy electron diffraction (LEED). LEEM allows for fast real-space imaging of the surface during growth of the FeO films, as well as selected-area diffraction measurements. Typical LEEM images were acquired at an electron energy (start voltage of the instrument) of 19~eV.

Single-crystal Ru substrates with (0001) orientation were cleaned by exposure to 5$\times10^{-8}$ Torr oxygen at 1000~K followed by flashing to 1500~K in vacuum (LEEM system) or by repeated cycles of 20 sec exposure to 2$\times10^{-7}$ Torr and flashing to 1500~K in vacuum (STM/XPS/LEED system).

The films were grown by O-MBE by depositing Fe from an 4mm-diameter Fe rod heated by electron bombardment in a variable background pressure of molecular oxygen. The Fe doser was calibrated by measuring the time needed to complete an Fe monolayer on Ru. (The density of the first Fe layer under our growth conditions is the same as the underlying Ru, i.e., the growth is pseudomorphic~\cite{santos_structure_2009}.) The Fe doses and oxygen coverages are given in ML$_{Fe}$, defined as the ratio of Fe or O atoms to Ru(0001) surface atoms.  Where appropriate, we will also talk about ML$_{FeO}$, corresponding to the density of an FeO(111) layer with an in-plane lattice spacing of 0.318~nm~\cite{monti_magnetism_2012}. Thus, 1~ML$_{Fe} \sim 1.38$ ML$_{FeO}$. The Fe flux was measured before and after the O-MBE experiments, and the change in the rate was less than 10\%. The rate used was $1.5\times10^{-3}$ ML/s.

\section{Results}

\subsection{Structural and Chemical Characterization of a FeO Bilayer Grown by O-MBE}
 
\begin{figure}
\centering{\includegraphics[width=0.7\linewidth]{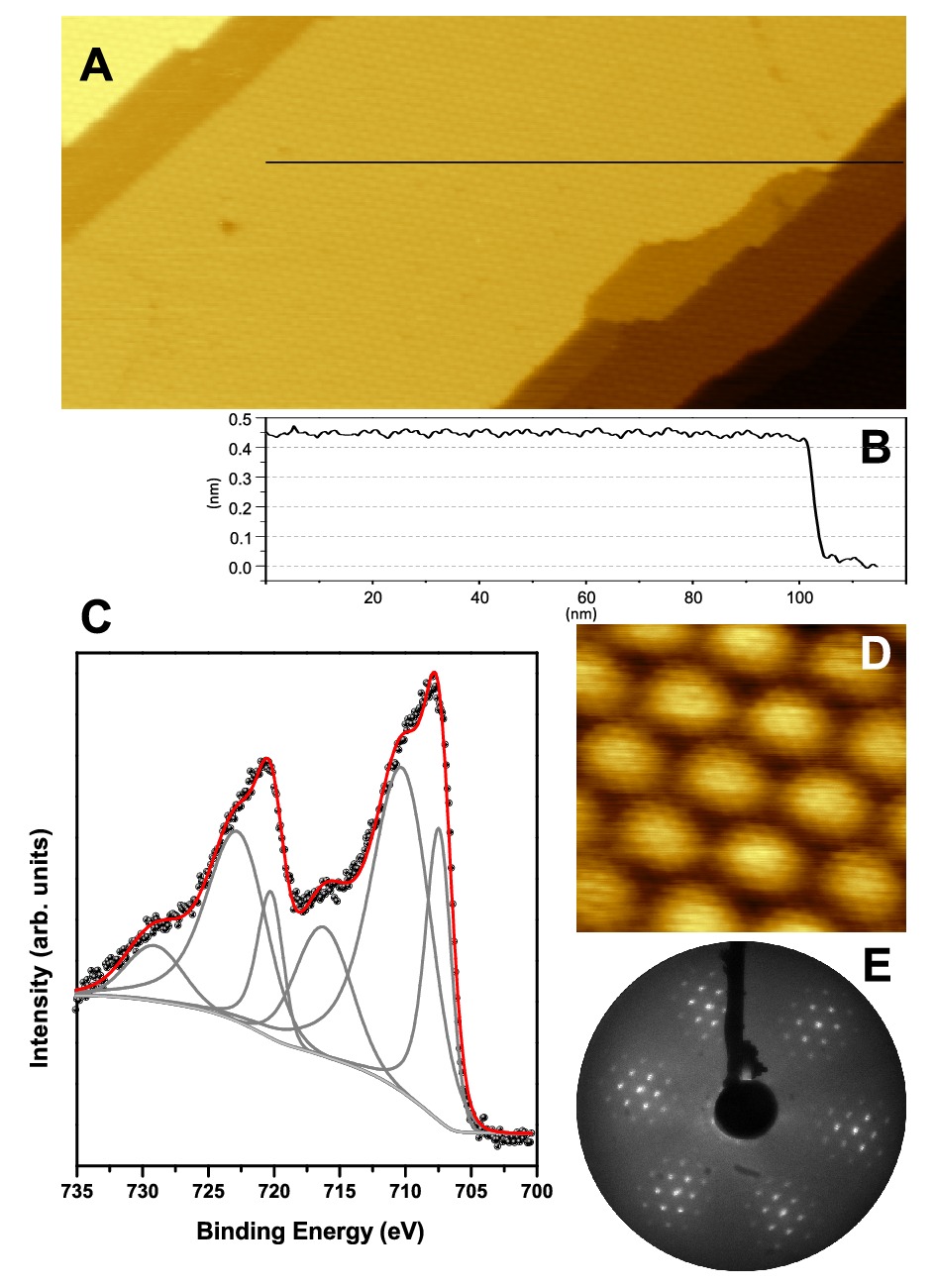}}
\caption{Characterization of FeO grown by O-MBE. A. STM image of an FeO bilayer exhibiting a moir\'{e} pattern from the misfit between the FeO and Ru lattices. Image size is 147~nm$\times$70~nm. B. profile along the black line in A. C. Fe 2$p$ XPS data. D. Atomic resolution STM image of the FeO bilayer. Image size is 6.8~nm$\times$6.3~nm. E. LEED at 60~eV. } 
\label{stm_and_xps}
\end{figure}

We begin by characterizing FeO films grown by depositing iron on a heated substrate in an oxygen background, O-MBE, and show that they are structurally and chemically equivalent to films grown by oxidizing a previously deposited iron layer. In Figure~\ref{stm_and_xps} we show STM, LEED and XPS data of a film grown by oxygen-assisted MBE by exposing the substrate at 900~K to an Fe flux of $1.5\times10^{-3}$ ML/s in $10^{-6}$~Torr of oxygen. The STM image (Figure~\ref{stm_and_xps}A) shows a nearly complete iron oxide film, which has a pronounced corrugation with an in-plane periodicity of $\sim$2~nm. The film thickness ($\sim 0.45$~nm) corresponds to two FeO layers (i.e., Fe-O-Fe-O) on top of the Ru substrate (Figure~\ref{stm_and_xps}B). Films grown by sequential steps of deposition and oxidation~\cite{ketteler_heteroepitaxial_2003} exhibit similar STM images except the thinner films are one Fe-O layer thick, not a double layer.  An atomically resolved image (Figure~\ref{stm_and_xps}D) shows protrusions separated by 0.32~nm, which corresponds to the in-plane atomic spacing of both monolayer and bilayer FeO films on Ru~\cite{ketteler_heteroepitaxial_2003}. The LEED pattern contains first-order spots whose separation corresponds to the atomic spacing of Figure~\ref{stm_and_xps}D and are consistent with the surface having an FeO structure. The satellite spots around the first-order spots are interpreted as arising from a superstructure that results from the coincidence of 6 FeO units over 7 Ru atoms. The Fe 2$p$ spectrum recorded from the grown FeO bilayer is quite complex (Figure~\ref{stm_and_xps}C) but agrees well with that reported for the surface of bulk FeO~\cite{McIntyre} or FeO grown on Pt~\cite{weiss_surface_2002}. It contains genuine photoemission peaks, multiplet splitting contributions, and shake-up satellite structure. A detailed description of the different contributions to the Fe 2$p$ core level spectrum of FeO can be found in Ref.~\cite{McIntyre}. In summary, all the techniques employed indicate that the FeO films grown by O-MBE are structurally and chemical the same as those grown by sequential steps of Fe deposition and oxidation.

\begin{figure}
\centering{\includegraphics[width=0.7\linewidth]{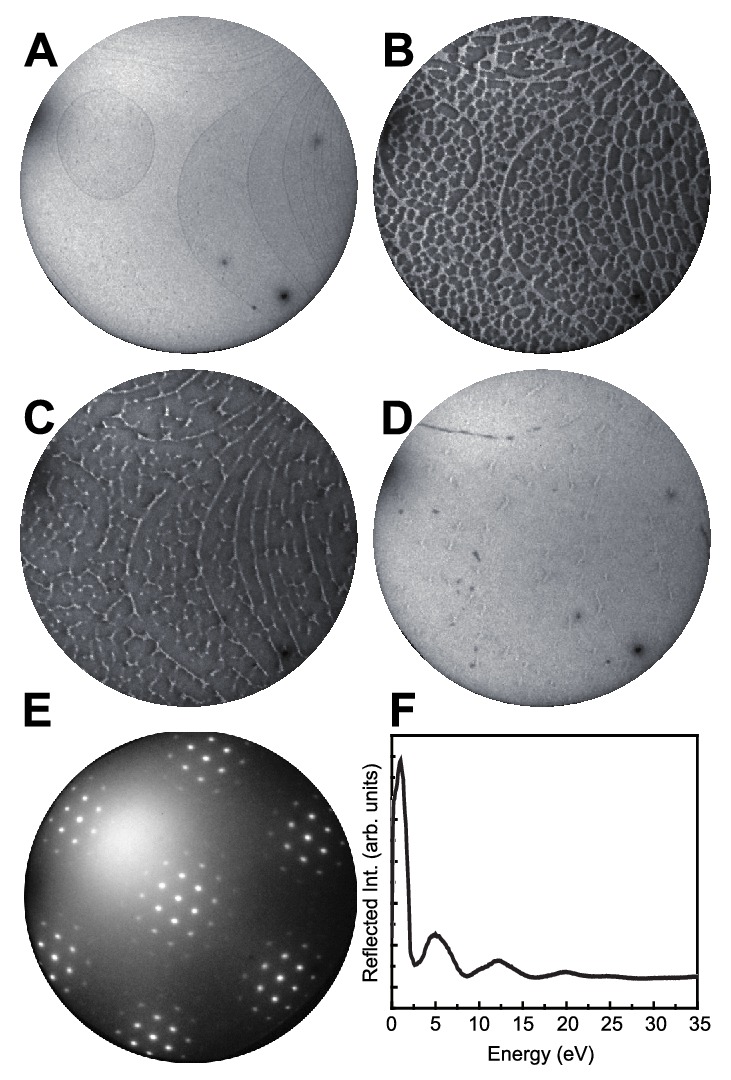}}
\caption{Growth of FeO in $10^{-6}$ Torr oxygen at 800~K. A-D. LEEM images extracted from a sequence acquired in real time during iron oxide growth (after +0,+354,+574 and +938 seconds, respectively). The image field of view is 10~$\mu$m. The electron energy is 19.4~eV. Total amount of iron deposited is 1.4~ML$_{Fe}$. E. LEED pattern of the final surface acquired using the LEEM instrument (42.3~eV). F. Electron reflectivity of the FeO surface.}
\label{LEEM_frames_highP}
\end{figure}

Figure~\ref{LEEM_frames_highP} shows LEEM images during O-MBE growth of FeO. Substrate steps are observed in the initial bare surface (see Figure~\ref{LEEM_frames_highP}A). When Fe deposition is started, iron oxide islands nucleate both on the substrate terraces and along the substrate steps (see Figure~\ref{LEEM_frames_highP}B). The islands grow in size until the surface gets completely covered by a continuous film of iron oxide (see Figure~\ref{LEEM_frames_highP}D). The LEED pattern obtained in the LEEM instrument after growth (Figure~\ref{LEEM_frames_highP}E) agrees well with the pattern from the conventional  diffractometer (Figure~\ref{stm_and_xps}E). Figure~\ref{LEEM_frames_highP}F shows how the electron reflectivity of the FeO surface (i.e., the intensity of the specular beam) varies with electron energy. (These measurements at low energy have been referred to as very-low energy electron diffraction, VLEED~\cite{PfnurSS1991}.) From the amount of iron deposited we estimate that the coverage in Figure~\ref{LEEM_frames_highP}D is a bilayer of FeO, i.e., Fe-O-Fe-O. As the images show only one stage of island nucleation followed by growth, we interpret that the islands initially nucleated under a pressure of 10$^{-6}$~Torr are of bilayer height (in agreement with the STM observation of Figure~\ref{stm_and_xps}). As we see below, single layer and bilayer FeO have different electron reflectivities, which enables ready differentiation in LEEM.

\subsection{Influence of Temperature and Oxygen Pressure} 

\begin{figure}
\centering{\includegraphics[width=0.6\linewidth]{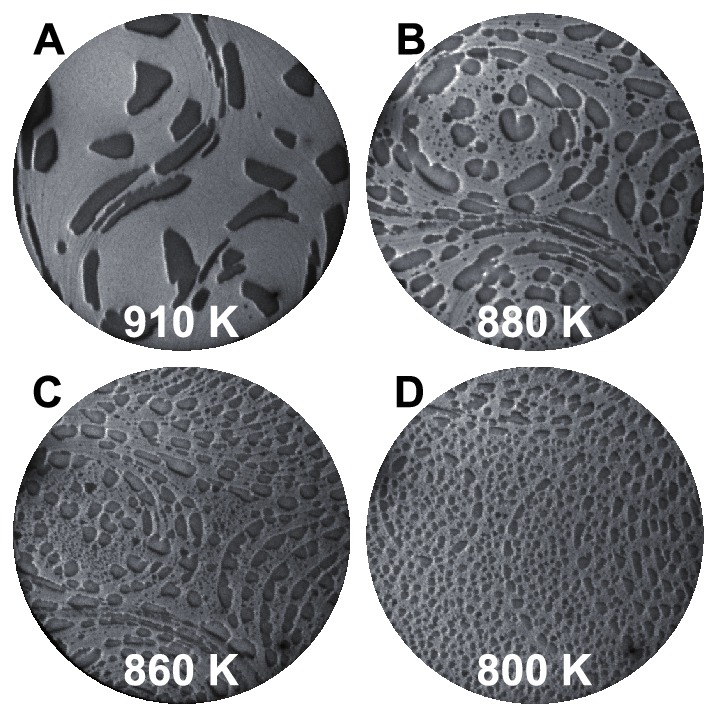}}
\caption{A-D. LEEM images acquired during FeO growth at the labeled substrate temperatures. Field of view 10~$\mu$m. The oxygen pressure is 10$^{-6}$~Torr. The electron energy is 19.4~eV.}
\label{islands_highP}
\end{figure}

Next we consider the effect of temperature and oxygen pressure on FeO growth by O-MBE. As mentioned above, Fe deposition in $10^{-6}$~Torr of oxygen leads to the nucleation and growth of bilayer-height islands, which eventually cover the substrate. This behavior is not affected by temperature in the range of 800--910~K. However, temperature strongly modifies the number of islands nucleated, as shown in Figure~\ref{islands_highP}, with higher temperature leading to fewer but larger islands. While at 910~K there are $3.4 \times 10^{7}$ islands/cm$^2$, at 800~K there are $1.3\times 10^{9}$ islands/cm$^2$. At the highest temperature, most islands nucleate at substrate step edges.

\begin{figure}
\centering{\includegraphics[width=0.6\linewidth]{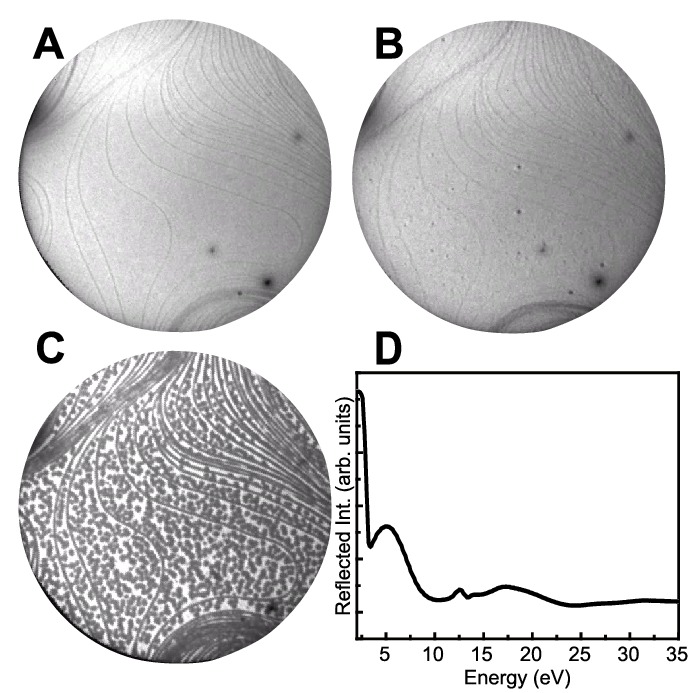}}
\caption{A-C. LEEM images of iron oxide film grown under an oxygen background pressure of $10^{-8}$~Torr. Total amount of iron deposited is 0.75~ML$_{Fe}$ (the time elapsed is +0, 262 and 514~seconds, respectively). Sample temperature is 800~K. Field of view is 10~$\mu$m and electron energy is 18.3~eV. D Reflectivity as a function of energy from the final film.}
\label{LEEM_frames_lowP}
\end{figure}

Oxygen pressure has a strong effect on both the island height and the nucleation density. We first discuss the height. Figure~\ref{LEEM_frames_lowP} shows growth at a lower oxygen background pressure, $10^{-8}$~Torr. At first glance, the image sequence is very similar to growth at the same temperature (800~K) but higher pressure (10$^{-6}$~Torr, Figure~\ref{LEEM_frames_highP}): islands nucleate and then grow until they cover the surface. But there is a crucial difference between them: depositing 1.4~ML$_{Fe}$ covers the surface with FeO at $10^{-6}$~Torr, but only $\sim$0.7~ML$_{Fe}$ is needed at $10^{-8}$~Torr. Furthermore, the electron reflectivity curves are quite different (compare Figure~\ref{LEEM_frames_highP}F and Figure~\ref{LEEM_frames_lowP}D), indicating the different nature of the two films. The LEED patterns are similar with the same satellite spots and lattice parameters. Considering the difference of lattice spacings of pseudomorphic iron and the iron oxide, the coverage for the complete film corresponds to $\sim$2~ML$_{FeO}$ and $\sim$1~ML$_{FeO}$ for $10^{-6}$~Torr and  $10^{-8}$~Torr, respectively. STM measurements (not shown) confirm that a complete film covering the substrate grown in $10^{-8}$~Torr is only one Fe-O layer thick. In contrast, when grown in $10^{-6}$~Torr it is two Fe-O layers thick (Fe-O-Fe-O). This difference in thickness explains the change in electron reflectivity as well as the difference in the amount of iron required to cover the surface with FeO.

\begin{figure}
\centering{\includegraphics[width=0.7\linewidth]{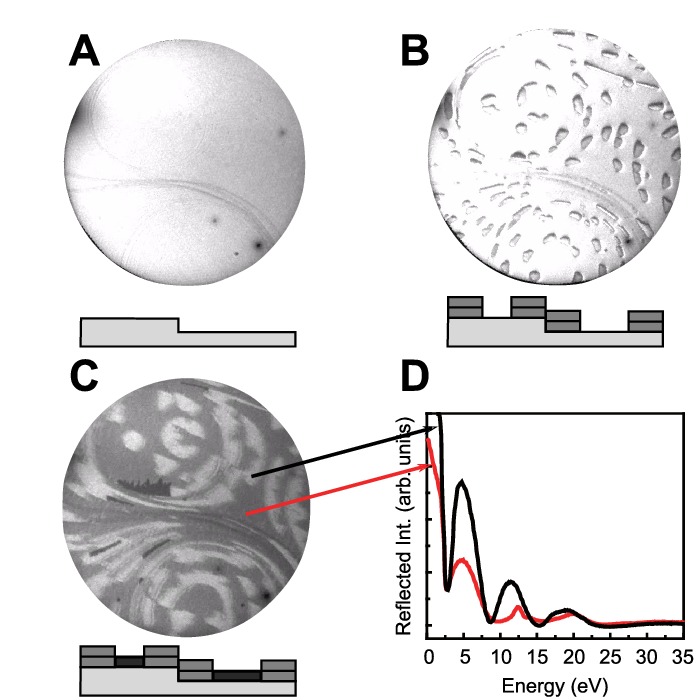}}
\caption{A-C. LEEM images of iron oxide film grown on Ru(0001) at 800~K in an oxygen background pressure of $10^{-7}$~Torr. Total amount of deposited iron is 1.1~ML$_{Fe}$. Field of View 10~$\mu$m and the electron energy is 18.0~eV. The time elapsed is 0, 120 and 744~seconds, respectively. The schematics below the images illustrate the cross-sectional morphology of the bare substrate (A), bilayer thick FeO (B), and monolayer plus bilayer thick FeO (C). D. Electron reflectivity as a function of energy from monolayer (red) and bilayer (black) regions of the final film.}
\label{LEEM_frames_medP}
\end{figure}

The intermediate pressure of $10^{-7}$~Torr produces a more complex film, as presented in Figure~\ref{LEEM_frames_medP}. The initially nucleated islands are all of bilayer height. However, as they grow, these bilayer islands switch to growing with monolayer height. The electron reflectivity, Figure~\ref{LEEM_frames_medP}D, identifies the areas that are FeO bilayers or monolayers. In the completed film, Figure~\ref{LEEM_frames_medP}C, the monolayer and bilayer regions are dark and medium grey, respectively.

\begin{figure}
\centering{\includegraphics[width=0.7\linewidth]{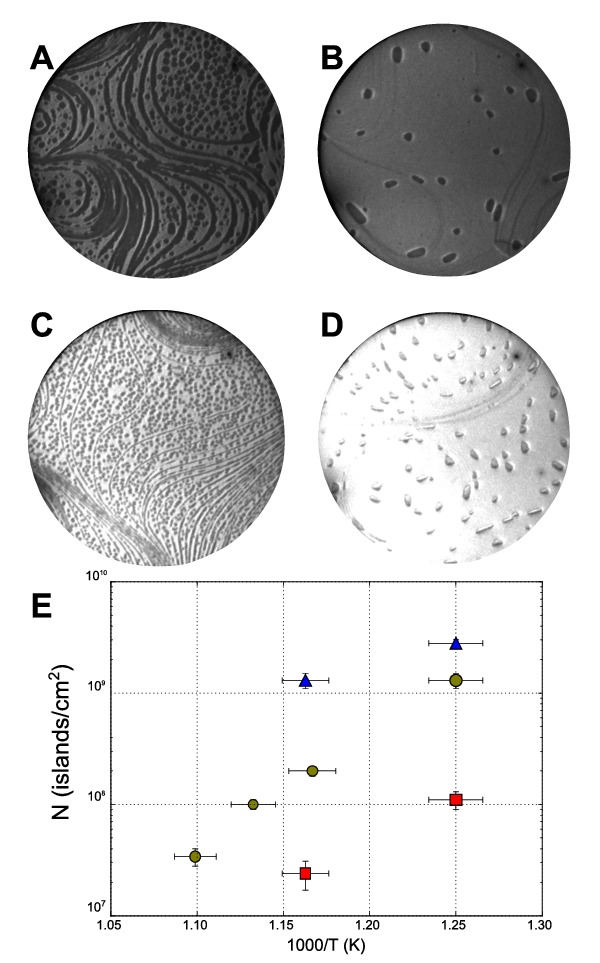}}
\caption{A--B. LEEM images acquired during the growth of FeO in $10^{-7}$~Torr at 860 and 800~K respectively. C--D. Same but under a pressure of $10^{-8}$~Torr of oxygen. The field of view is 10~$\mu$m. E Plot of the island densities vs. inverse temperature. The symbols correspond to the data at the different pressures: green circles for $10^{-6}$~Torr, red squares for $10^{-7}$~Torr and blue triangles for $10^{-8}$~Torr.}
\label{islands_medP}
\label{islands_lowP}
\end{figure}

The effect of temperature on the number of islands nucleated is presented in Figure~\ref{islands_medP} for $10^{-7}$~Torr and $10^{-8}$~Torr. The trend is similar to the previously presented data at the higher pressure ($10^{-6} $~Torr, Figure~\ref{islands_highP}): the island density decreases with increasing temperature. At a given temperature the island density is the lowest at the intermediate pressure of $10^{-7}$~Torr.

In summary, at $10^{-6}$ Torr, FeO always grows as a bilayer. At $10^{-7}$~Torr, the initial islands are bilayer. However, as the islands expand, they switch and grow as monolayer FeO. Finally, at $10^{-8}$~Torr, only monolayer FeO grows. The number of islands nucleated decreases with temperature and the influence of oxygen pressure is non-monotonic: the fewest islands nucleate at an oxygen pressure of $10^{-7}$~Torr.

\subsection{FeO growth with a limited supply of oxygen}
\label{sec:limited}

\begin{figure}
\centering{\includegraphics[width=0.5\linewidth]{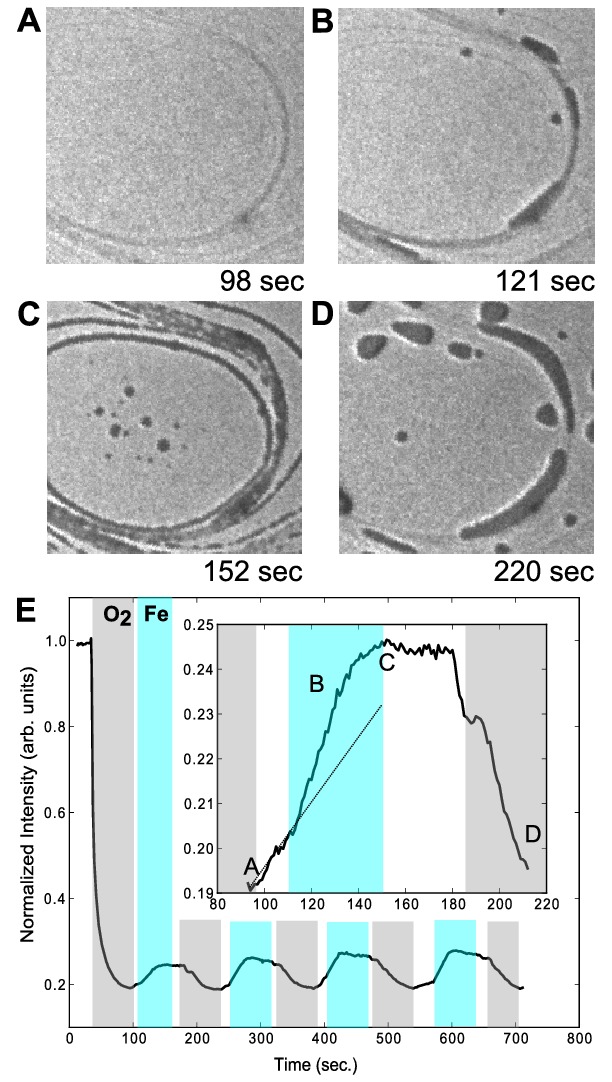}}
\caption{A-D. LEEM images acquired at different stages (indicated by the time when each frame was acquired) of iron oxide growth performed in separate steps of oxygen saturation and iron deposition. The field of view is 10~$\mu$m.  The electron energy is 19~eV. E. Electron reflectivity vs time from the bare areas of the substrate (i.e., areas without iron oxide islands). The grey/blue colored regions indicate when oxygen/iron were being dosed, respectively. The inset is an enlargement of the initial data where the labels refer to the times when the correspoding LEEM snapshots were acquired.}
\label{oxygen_limited_growth}
\end{figure}

Additional insight about the FeO growth mechanism can be gained by limiting the supply of oxygen during growth, as we next describe. Instead of a constant background pressure of oxygen, iron oxide was grown at 800~K on a surface saturated with oxygen by exposure to 6.5~L (1 Langmuir = 10$^{-6}$ Torr$\times$sec) at a pressure of $10^{-7}$ Torr (Figure~\ref{oxygen_limited_growth}A). Iron was then dosed on the surface at the same rate as in the previous experiments. This procedure led to the nucleation of bilayer FeO islands, the dark regions in Figure~\ref{oxygen_limited_growth}B, as confirmed from post-growth reflectivity measurements. Thus oxygen adsorbed on the Ru reacts with the deposited Fe to nucleate FeO islands. Initially these islands grew while depositing additional iron. Then they stopped growing. At this time the substrate steps became decorated and new islands nucleated in the middle of the Ru terraces (Figure~\ref{oxygen_limited_growth}C). The interpretation is that eventually there is not enough oxygen available on the ruthenium surface to support continued growth of stoichiometric FeO. Then iron-oxygen islands with different composition grow. Regardless of their particular composition, the new islands display a markedly different growth behavior. As soon as the change in growth mode was detected, the iron doser was stopped. Then the surface was saturated again with oxygen by dosing an additional 6.5~L. The iron-rich islands in the center of the image mostly vanish (Figure~\ref{oxygen_limited_growth}D), leading mostly to the growth of the initial FeO islands and the nucleation of few additional islands.


Figure~\ref{oxygen_limited_growth}E shows the electron reflectivity from the bare substrate regions during cycles of oxygen dosing (grey shaded regions in the plot) followed by Fe dosing (blue shaded regions). Changes in reflectivity can be related to the oxygen concentration on the ruthenium~\cite{FigueraSS2006,LoginovaNJP2008}. The saturation of the surface upon oxygen exposure is detected by an exponential decrease of the reflected intensity as the steady-state oxygen coverage at 800~K is reached after $\sim$60 seconds. When the oxygen background is removed, the reflected intensity increases linearly with time (see the inset of Figure~\ref{oxygen_limited_growth}E). We attribute this increase to oxygen being removed from the surface, either by desorption or by dissolution into the bulk of the crystal~\cite{ErtlPRL1996}. The slope of the reflected intensity vs. time increases during Fe deposition, which indicates that oxygen is being removed faster. This can be rationalized by assuming that some fraction of the adsorbed oxygen is being incorporated into the growing FeO islands. When the reflected electron intensity stops changing (point C in the insert to Figure~\ref{oxygen_limited_growth}E)), distinct islands (center of Figure~\ref{oxygen_limited_growth}C) nucleate on the substrate. Because of their distinct growth behavior, we interpret these new islands as oxygen-deficient FeO. Stopping the iron dosing and exposing to oxygen replenishes the density of the oxygen on the Ru, as show by the substrate's electron reflectivity (point D in the inset of Figure~\ref{oxygen_limited_growth}E) recovering nearly to the value observed at the end of the first oxygen dose (point A in the insert to Figure~\ref{oxygen_limited_growth}E). More importantly, the smaller, oxygen-deficient islands disappear and the original FeO islands grow. Repetitions of the cycle lead to similar behavior and a net growth of the FeO islands.

\section{Discussion}

We find that the FeO phase grown by oxygen-assisted MBE is structurally and chemically identical to the oxide produced by depositing and then oxidizing Fe. But the O-MBE technique produced an unexpected result: the morphology of the film changes from {\em bilayer} islands that coalesce to form a continuous film at $10^{-6}$~Torr to {\em monolayer} islands that grow into a continuous film at $10^{-8}$~Torr. And during growth at an intermediate pressure, $10^{-7}$~Torr, the film initially grew as a bilayer but then switched to growing as a monolayer. These results are surprising because the monolayer and bilayer films have the same composition and structure. So changing the background oxygen pressure does not change the stoichiometry of the oxide. Instead it changes the film morphology (thickness). We next propose an explanation for this striking effect of oxygen pressure. 

In O-MBE, the FeO-free Ru substrate is covered by adsorbed oxygen. Figure~\ref{oxygen_limited_growth} (Section~\ref{sec:limited}) shows that FeO grows from adsorbed oxygen reacting with the deposited Fe. The key to understanding the effect of oxygen pressure on FeO thickness comes from comparing the areal density of the adsorbed oxygen with the oxygen density in monolayer and bilayer FeO, respectively, as we next explain. The areal density of oxygen in an FeO monolayer is a factor of 0.72 less than the density of Ru atoms in the surface layer. If the adsorbed oxygen concentration is lower than 0.72 ML$_{Fe}$, forming monolayer islands of FeO reduces the oxygen density on the ruthenium. On the contrary, if the oxygen density is higher than 0.72 ML$_{Fe}$, then forming monolayer FeO  {\em increases} the oxygen density. But this increase in density is hindered. First, there is a limit to the density of oxygen on ruthenium: one oxygen atom per ruthenium atom. Second, even for lower densities oxygen-oxygen interactions on Ru are repulsive, as shown by the decrease of the binding energy as a function of coverage~\cite{ErtlPRL1996}. Thus, we propose that if the oxygen density is sufficiently high, the growth of monolayer FeO is hindered because this increases the density of adsorbed oxygen. Instead, FeO bilayers, with an oxygen density of 1.44 ML$_{Fe}$ (due to the two oxygen planes of the bilayer), grow and remove adsorbed oxygen.

So what evidence supports this mechanism? Data from two methods show that the concentration of adsorbed oxygen is near the value (0.72 ML) proposed to select either monolayer or bilayer growth. First, after cooling to room temperature, FeO islands grown in $10^{-6}$ Torr are surrounded by oxygen-covered Ru with a $2\times2$ structure, as evidenced by STM. An oxygen concentration above about 0.75 ML gives a $2\times2$ structure labeled as $(2\times2)-3$O~\cite{kostov_observation_1997}. Second, a rough estimate from the reflectivity changes during FeO growth (see \ref{ap:limited}) gives a similar oxygen density. We do note, however, that oxygen densities near 0.75 ML seem high for our continuous or sequential exposures to $10^{-7}$ Torr oxygen, for which we estimate doses in the range of 10~L~\cite{PfnurSS1991,ErtlPRL1996,MadeySS1975,PiercyPRB1992,lizzitPRB2001}.

In $10^{-7}$ Torr, bilayer FeO islands grow initially. But monolayer FeO grows later at this pressure (see Figure~\ref{LEEM_frames_medP}). This can be rationalized by decreased oxygen concentration on the ruthenium caused by the smaller sticking coefficient of oxygen on FeO vs. ruthenium. This difference in sticking coefficient suggests that maintaining the oxygen concentration required to complete a uniform bilayer may be difficult at lower oxygen pressures.

\begin{figure}
\centering{\includegraphics[width=0.6\linewidth]{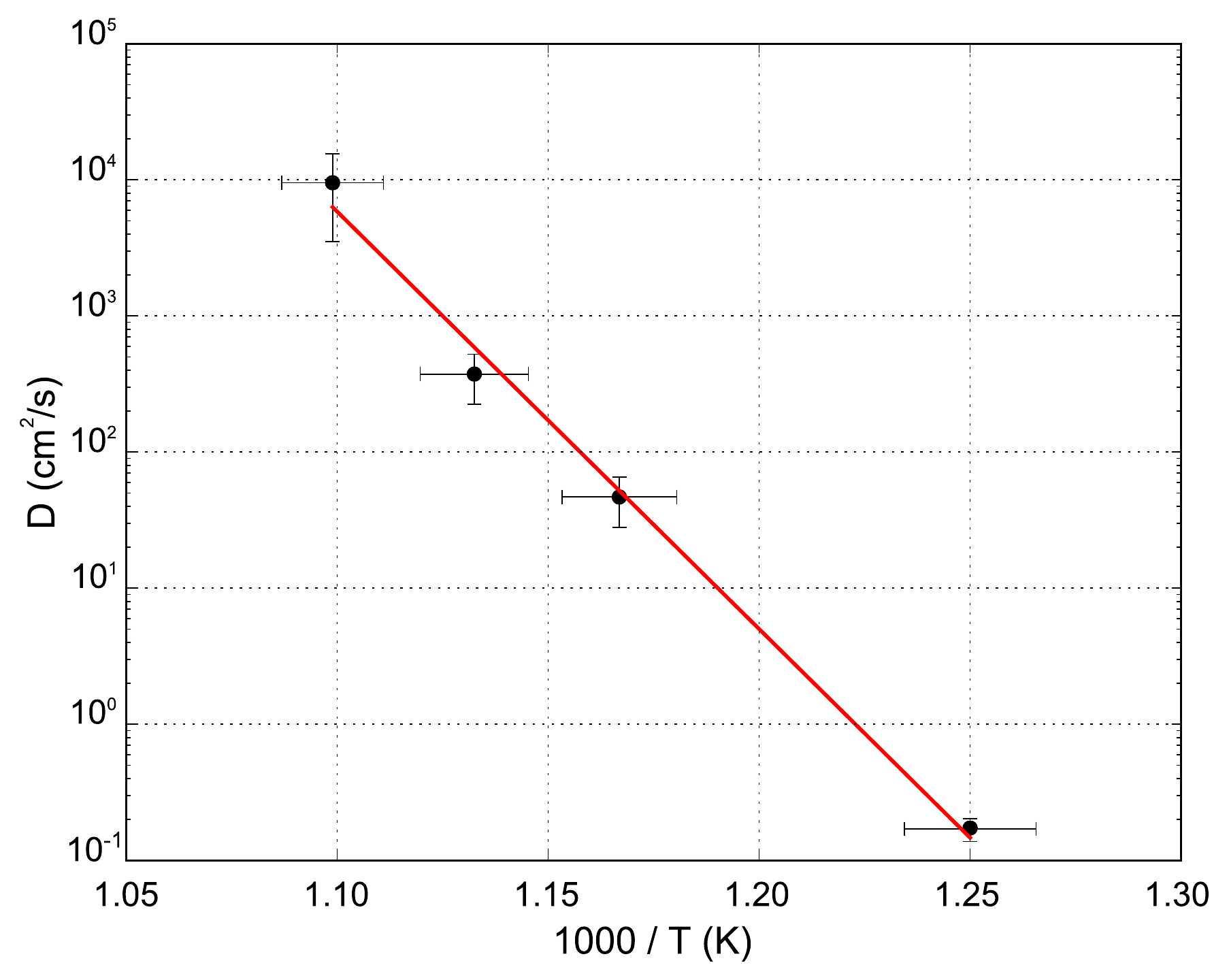}}
\caption{Plot of the surface diffusion vs. inverse temperature, as extracted from a simple nucleation model (see text) and the nucleation data corresponding to an oxygen pressure of 10$^{-6}$~Torr.}
\label{plot_diffusion}
\end{figure}

The density of FeO islands as a function of temperature and pressure is also striking. While at a given pressure it always decreases with temperature, pressure plays a role, as readily detected in Figure~\ref{islands_lowP}. On one hand, the island densities are lower at the intermediate pressure ($10^{-7}$~Torr). On the other, the slope in the logarithmic plot of island density vs. inverse temperature is more similar for  $10^{-7}$ and $10^{-6}$~Torr than for $10^{-8}$~Torr. This latter effect might be related to the bilayer vs. monolayer growth regimes. Further study is required to understand this complex behavior.

Meanwhile, we can interpret the results at a given oxygen pressure in terms of the simple nucleation and growth models that have explained the density decrease of metal and semiconductor islands with increasing temperature~\cite{VenablesRPP1984,MoSS1992,StroscioPRL1993}. As the iron atoms arrive at the surface, they probably form some oxygen-iron species that diffuse on the surface before reaching either an already nucleated FeO island or another oxygen-iron species. In the former case, the island grows. In the latter case, a new island nucleates. This leads to two regimes~\cite{VenablesRPP1984}: the initial nucleation regime, where the number of islands keeps increasing, followed by the growth regime, where the already nucleated islands keep growing. In this simple picture, increasing the temperature causes faster surface diffusion of the Fe-O complexes, allowing them to explore larger areas, which results in fewer nucleated islands. 

But we note that even in homoepitaxial, single-component growth this model can be too simple. Much more so for our heteroepitaxial, bi-component oxide growth. Nevertheless, it is instructive to estimate the main energy barrier involved by assuming the simplest nucleation model~\cite{MoSS1992} for the highest pressure experiments. Then the  island area is disregarded (as well as the bilayer thickness) and the critical nuclei is assumed to be a dimer. The relationship between the island density and the diffusion coefficient is then  $N^3\sim R\theta/D_S$ where $N$ is the island density, $R$ is the rate of arrival of iron atoms, $\theta$ is the coverage and $D_{S}$ the diffusion coefficient. Solving $D_S$ using the experimentally measured islands density gives the plot of Figure~\ref{plot_diffusion}. Assuming that the diffusion of the iron (or iron-oxygen complex) on the surface follows an Arrhenius form $D_S=D_{S_0}\exp (-E_S/kT)$, where $D_{S_{0}}$ is a pre-factor and $E_{S}$ is the diffusion barrier (eV), $T$ the temperature and $k$ the Boltzmann constant, the dependency of $D_S$ should follow the line shown in Figure~\ref{plot_diffusion}. It 
corresponds to a diffusion barrier of $(5.9\pm0.5)$~eV. This number is much 
larger than the typical diffusion barriers for surface diffusion on metals, but it is in the expected range for surface diffusion on oxides~\cite{RobertsonJNM1969}. Nevertheless we warn that the model is too simplistic and it cannot be expected to capture the detailed growth process, as reflected by an unrealistic diffusion prefactor of $10^{37}$ cm$^{2}/$sec. As directly observed in Figure~\ref{islands_lowP}E, the estimated energy barrier is much smaller (smaller slope) for the lowest pressure of $10^{-8}$ (2.2~eV), while the intermediate pressure value is closer to the high pressure case (4.5~eV).  

\section{Summary}

We have studied the initial stages of FeO growth on Ru(0001) by the simultaneous deposition of oxygen and iron at elevated temperatures. In an excess of oxygen (in a background pressure of $10^{-6}$~Torr), FeO grows by the nucleation and spreading of bilayer-thick islands, which eventually coalesce into a complete layer. There is a strong influence of oxygen pressure: at 100 times lower oxygen pressure, only monolayer FeO grows. At intermediate pressures, the initial oxide is bilayer but eventually monolayer FeO grows. We explain the influence of oxygen pressure by considering how the concentration of oxygen adsorbed on the Ru changes as this oxygen is incorporated into the film. Monolayer FeO formation can either decrease or increase the density of adsorbed oxygen. The latter case, which occurs at high concentrations of adsorbed oxygen, increases the concentration above a critical density. This suppresses monolayer growth, leading to exclusive bilayer growth. Increasing the substrate temperature decreases the island density. But the evolution of island density as a function of oxygen pressure is not monotonic, underlying the complexity of the FeO system.

\ack This research was supported by the Spanish Ministry of Education and Science under Projects No.~MAT2009-14578-C03-01 and MAT2009-14578-C03-02, and by the U.S. Department of Energy, Office of Basic Energy Sciences, Division of Materials Sciences and Engineering, under contract No. DE-AC04-94AL85000. IP and MM thank the Spanish Ministry of Science and Innovation for support through FPI fellowships.

\appendix
\section{Estimate of concentration from the limited oxygen experiment}
\label{ap:limited}

We use changes in electron reflectivity to estimate the density of oxygen adsorbed on the ruthenium during FeO growth. At the electron beam energy used for imaging in Figure~\ref{oxygen_limited_growth} (19~eV), the reflectivity from the clean Ru is relatively high because a band gap in Ru(0001) results in a low density of unoccupied states. Adsorbed oxygen atoms provide additional scattering channels, which decrease the reflectivity. The changes in electron reflectivity for O/Ru have been studied as an structural method by Pfn\"ur et al.~\cite{PfnurSS1991}. The changes can also be used to track thermal adatom concentrations, as reported for Au~\cite{FigueraSS2006} and C~\cite{LoginovaNJP2008}.



Here we estimate the amount of oxygen adsorbed on Ru just before iron deposition is started in the sequential dosing experiment described in Section~\ref{sec:limited} (see Figure~\ref{oxygen_limited_growth}). Disregarding the removal of oxygen due to other factors unrelated to the FeO growth (such as desorption or dissolution into the Ru), mass conservation dictates that:
\[
 c_i = \theta_{FeO} c_{FeO} + (1-\theta_{FeO}) c_f
\]
  where $\theta_{FeO}$ is the fraction of the surface covered by the islands of FeO at the end of the cycle, $c_{FeO}$ the density of oxygen in those islands, and $c_i, c_f$ are the densities of adsorbed oxygen on the ruthenium substrate before and after the growth of the FeO islands, respectively. We refer to all the oxygen densities relative to the ruthenium substrate (i.e., a density of 1 corresponds to one atom per Ru substrate atom). The coverage of FeO islands is obtained from the amount of iron deposited on the surface. The initial concentration can then be written in terms of the ratio $r=\frac{c_i}{c_f}$:
\[
 c_i = \frac{\theta_{FeO} c_{FeO}}{1-\frac{1-\theta_{FeO}}{r}}
\]
We assume that the changes in electron reflectivity are proportional to the changes in the adsorbed oxygen density [$c=\alpha (1-i)$ where $i$ is the local reflected intensity].  (The dependence of oxygen density on reflectivity is, though, unlikely to be linear for our large range of concentrations.) Then the ratio $r$ can be estimated without knowing the proportionality constant $\alpha$. From the observed ratio between intensities in the first growth step of Figure~\ref{oxygen_limited_growth}E, $\frac{c_i}{c_f}\sim 1.016$, and FeO coverage (estimated from the deposited iron) during the first deposition cycle ($0.017$ ML$_{2FeO}$ in bilayer form) we obtain an 
initial oxygen density in that experiment of $c_i\sim 0.75$ ML$_{Ru}$. 

\section*{References}
\bibliographystyle{unsrt}
\bibliography{FeO_next}

\end{document}